\documentclass[aps,pra,twocolumn,showpacs]{revtex4}
\usepackage{amsfonts}

\usepackage{amsmath}
\usepackage{amssymb}
\usepackage{graphicx}
\begin{document}

\title{Test for entanglement using physically \\
observable witness operators and positive maps}
\author{Kai Chen}
\email{kchen@aphy.iphy.ac.cn}
\author{Ling-An Wu}
\email{wula@aphy.iphy.ac.cn}
\affiliation{Laboratory of Optical Physics, Institute of Physics, Chinese Academy of
Sciences, Beijing 100080, P.R. China}

\begin{abstract}
Motivated by the Peres-Horodecki criterion and the realignment criterion we
develop a more powerful method to identify entangled states for any
bipartite system through a universal construction of the witness operator.
The method also gives a new family of positive but non-completely positive
maps of arbitrary high dimensions, which provide a much better test than the
witness operators themselves. Moreover, we find that there are two types of
positive maps that can detect $2\times N$ and $4\times N$ bound entangled
states. Since entanglement witnesses are physical observables and may be
measured locally our construction could be of great significance for future
experiments.
\end{abstract}

\date{\today}
\maketitle


\section{Introduction}

Quantum entangled states lie at the heart of the rapidly developing field of
quantum information science, which encompasses important potential
applications such as quantum communication, quantum computation and quantum
lithography  \cite{pre98,nielsen,zeilinger}. However, the fundamental nature
of entangled states has tantalized physicists since the earliest days of
quantum mechanics, and even today is by no means fully understood.

One of the most basic problems is that how can one tell if a quantum state
is entangled?, and how entangled is it still after some noisy quantum
process (e.g. long distance quantum communication)?

A pure entangled state is a quantum state which cannot be factorized, i.e., $%
\left\vert \Psi \right\rangle _{AB}\neq \left\vert \psi \right\rangle
_{A}\left\vert \phi \right\rangle _{B}$, and shows remarkable nonlocal
quantum correlations. From a practical point of view, the state of a
composite quantum system which usually becomes a mixed state after a noisy
process, is called \emph{unentangled} or \emph{separable} if it can be
prepared in a \textquotedblleft local\textquotedblright\ or
\textquotedblleft classical\textquotedblright\ way. It can then be expressed
as an ensemble realization of pure product states $\left\vert \psi
_{i}\right\rangle _{A}\left\vert \phi _{i}\right\rangle _{B}$ occurring with
a certain probability $p_{i}$ and the density matrix $\rho
_{AB}=\sum_{i}p_{i}\rho _{i}^{A}\otimes \rho _{i}^{B},$ where $\rho
_{i}^{A}=\left\vert \psi _{i}\right\rangle _{A}\left\langle \psi
_{i}\right\vert $, $\rho _{i}^{B}=\left\vert \phi _{i}\right\rangle
_{B}\left\langle \phi _{i}\right\vert $, $\sum_{i}p_{i}=1$, and $\left\vert
\psi _{i}\right\rangle _{A}$, $\left\vert \phi _{i}\right\rangle _{B}$ are
normalized pure states of subsystems $A$ and $B$, respectively \cite%
{werner89}. If no convex linear combination of $\rho _{i}^{A}\otimes \rho
_{i}^{B}$ exists for a given $\rho _{AB}$, then the state is called
\textquotedblleft entangled\textquotedblright .

However, for a generic mixed state $\rho _{AB}$, finding a separable
decomposition or proving that it does not exist is a non-trivial task (see
the recent good reviews \cite{lbck00,terhal01,3hreview,bruss01} and
references therein). There have been many efforts in recent years to analyze
the separability and quantitative character of quantum entanglement. The
Bell inequalities satisfied by a separable system give the first necessary
condition for separability \cite{Bell64}. In 1996, Peres made an important
step towards proving that, for a separable state, the partial transposition
with respect to one subsystem of a bipartite density matrix is positive, $%
\rho ^{T_{A}}\geq 0$. This is known as the PPT (positive partial
transposition) criterion or Peres-Horodecki criterion \cite{peres}. By
establishing a close connection between positive map theory and
separability, Horodecki \textit{et al.} promptly showed that this is a
sufficient condition for separability for bipartite systems of $2\times 2$
and $2\times 3$ \cite{3hPLA223}. Regarding the quantitative character of
entanglement, Wootters succeeded in giving an elegant formula to compute the
\textquotedblleft \textit{entanglement of formation}\textquotedblright\ \cite%
{be96} of $2\times 2$ mixtures, thus giving also a separability criterion
\cite{wo98}.

Very recently, Rudolph and other authors \cite%
{ru02,ChenQIC03,chenPLA02,rupra} proposed a new operational criterion for
separability, the \emph{realignment} criterion (named, thus, following the
suggestion of Ref. \cite{Horo02}) which is equivalent to the \emph{%
computational cross norm} criterion of Ref. \cite{ru02}). The criterion is
very simple to apply and shows a dramatic ability to detect bound entangled
states (BESs) \cite{hPLA97} in any high dimension \cite{ChenQIC03}. It is
even strong enough to detect the true tripartite entanglement shown in Ref.
\cite{Horo02}.

An alternative method to detect entanglement is to construct so-called
entanglement witnesses (EWs) \cite{3hPLA223,terhal00,lkch00} and positive
maps (PMs) \cite{pmaps}. Entanglement witnesses \cite{3hPLA223,terhal00} are
physical observables that can \textquotedblleft detect\textquotedblright\
the presence of entanglement. Starting from the witness operators one can
also obtain PMs \cite{Jami72} that detect more entanglement. Although there
are constructions of EWs related to the unextendible bases of product
vectors in Ref. \cite{terhal00} and to the existence of \textquotedblleft
edge\textquotedblright\ positive partial transpose entangled states (PPTES)
in Ref. \cite{lkch00}, a universal construction of EWs and PMs for a general
bipartite quantum state has still to be discovered.

The aim of this paper is to introduce a new powerful technique for universal
construction of EWs and PMs for any bipartite density matrix and to obtain a
stronger operational test for identifying entanglement. Our starting point
will be the $PPT$ criterion and the realignment criterion for separability.
The universal construction is given in Sec. \ref{sec2}, and several typical
examples of entangled states which can be recognized by the corresponding
EWs and PMs are presented in Sec. \ref{sec3}. We show that many of the
recognized bound entangled states cannot be detected by the realignment
criterion or any of the EWs and PMs constructed previously in the
literature. Moreover, we demonstrate in Sec. \ref{sec4} that there are two
types of positive maps that can detect systematically the $2\times N$ and $%
4\times N$ bound entangled states. A brief summary and discussion are given
in the last section.

\section{Universal construction of entangled witnesses and positive maps for
identifying entanglement}

\label{sec2}

In this section we will give two universal constructions for EWs and PMs for
any bipartite density matrix that provide stronger operational tests for
identifying entanglement. The starting points are the $PPT$ criterion \cite%
{peres,3hPLA223} and the realignment criterion for separability \cite%
{ru02,ChenQIC03,chenPLA02,rupra}. The main tools that we shall use are drawn
from the general theory of matrix analysis \cite{hornt1,hornt}.

\subsection{Some notation}

\label{sec2.1} The various matrix operations from \cite{hornt1,hornt} that
we need will employ the following notation.

\noindent \textbf{Definition:} \emph{For each $m\times n$ matrix $A=[a_{ij}]$%
, where $a_{ij}$ is the matrix entry of A, we define the vector $vec(A)$ as}
\begin{equation*}
vec(A)=[a_{11},\cdots ,a_{m1},a_{12},\cdots ,a_{m2},\cdots ,a_{1n},\cdots
,a_{mn}]^{T}.
\end{equation*}%
Here the superscript \textquotedblleft $T$" means standard transposition.
Let $Z$ be an $m\times m$ block matrix with block size $n\times n$. We
define the following \textquotedblleft realignment\textquotedblright\
operation $\mathcal{R}$ to change $Z$ to a realigned matrix $\widetilde{Z}$
of size $m^{2}\times n^{2}$ that contains the same elements as $Z$ but in
different positions as follows:
\begin{equation}
\mathcal{R}(Z)\equiv \widetilde{Z}\equiv \left[
\begin{array}{c}
vec(Z_{1,1})^{T} \\
\vdots  \\
vec(Z_{m,1})^{T} \\
\vdots  \\
vec(Z_{1,m})^{T} \\
\vdots  \\
vec(Z_{m,m})^{T}%
\end{array}%
\right] .  \label{realign}
\end{equation}

For example, a $2\times 2$ bipartite density matrix $\rho $ can be
transformed as
\begin{align}
\rho & =\left(
\begin{array}{cc|cc}
\rho _{11} & \rho _{12} & \rho _{13} & \rho _{14} \\
\rho _{21} & \rho _{22} & \rho _{23} & \rho _{24} \\ \hline
\rho _{31} & \rho _{32} & \rho _{33} & \rho _{34} \\
\rho _{41} & \rho _{42} & \rho _{43} & \rho _{44}%
\end{array}%
\right)   \notag \\
& \longrightarrow \mathcal{R}(\rho )=\left(
\begin{array}{cccc}
\rho _{11} & \rho _{21} & \rho _{12} & \rho _{22} \\ \hline
\rho _{31} & \rho _{41} & \rho _{32} & \rho _{42} \\ \hline
\rho _{13} & \rho _{23} & \rho _{14} & \rho _{24} \\ \hline
\rho _{33} & \rho _{43} & \rho _{34} & \rho _{44}%
\end{array}%
\right) .
\end{align}

\subsection{The realignment criterion}

\label{sec2.2}

Motivated by the Kronecker product approximation technique for a matrix \cite%
{loan,pits}, we developed a very simple method to obtain the realignment
criterion in Ref. \cite{ChenQIC03} (called the cross norm criterion in \cite%
{ru02}). To recollect, the criterion says that, \emph{for any separable $%
m\times n$ bipartite density matrix $\rho _{AB},$ the $m^{2}\times n^{2}$
matrix $\mathcal{R}(\rho _{AB})$ should satisfy $||\mathcal{R}(\rho
_{AB})||\leq 1,$ where $||\mathcal{\cdot }||$ means the trace norm defined
as $||G||=Tr((GG^{\dagger })^{1/2})$. Thus $||\mathcal{R}(\rho _{AB})||>1$
implies the presence of entanglement in $\rho _{AB}.$}

This criterion is strong enough to detect most of the bound entangled states
in the literature, as shown in Ref. \cite{ChenQIC03}, and holds even for
genuine multipartite systems, as shown in Ref. \cite{Horo02}.

\subsection{Entanglement witnesses and positive maps}

\label{sec2.3}

\noindent \textbf{Entanglement witness:} an entanglement witness is a
Hermitian operator $W=W^{\dagger }$ acting on the Hilbert space $\mathcal{H}=%
\mathcal{H}_{A}\otimes \mathcal{H}_{B}$ that satisfies $Tr(W\rho _{A}\otimes
\rho _{B})\geq 0$ for any pure separable state $\rho _{A}\otimes \rho _{B}$,
and has at least one negative eigenvalue. If a density matrix $\rho $
satisfies $Tr(W\rho )<0$, then $\rho $ is an entangled state and we say that
$W$ \textquotedblleft detects\textquotedblright\ entanglement in $\rho $
\cite{3hPLA223,terhal00,lkch00}. It has been shown in Ref. \cite{3hPLA223}
that a given density matrix is entangled if and only if there exists an EW
that detects it.

\noindent \textbf{Positive map:} it was shown in Ref. \cite{3hPLA223} that $%
\rho $ is separable iff for any positive map $\Lambda $ the inequality
\begin{equation}
(Id_{A}\otimes \Lambda )\rho \geq 0  \label{posimap}
\end{equation}%
holds where $Id_{A}$ means a identity matrix with respect to the A
subsystem. In practice, detecting entanglement only involves finding those
maps which are positive but \emph{not} completely positive (non-CP), since a
CP map will satisfy Eq. (\ref{posimap}) for any given separable $\rho $ \cite%
{3hPLA223}.

It was shown in \cite{Jami72} that there is a close connection between a
positive map and the entanglement witness, i.e., the \emph{Jamio\l kowski
isomorphism}
\begin{equation}
W=(Id_{A}\otimes\Lambda)P_{+}^{m},   \label{isomor}
\end{equation}
where $P_{+}^{m}=|\Phi\rangle\langle\Phi|$ and $|\Phi\rangle=\frac{1}{\sqrt {%
m}}\sum_{i=1}^{m}|\,ii\rangle$ is the maximally entangled state in $\mathcal{%
H}_{A}\otimes\mathcal{H}_{A}$.

\subsection{Universal construction of EWs}

\label{sec2.4}

With the above mentioned notation and concepts in mind we will now derive
the main result of this paper: two universal constructions for EWs and PMs
to identify entanglement of bipartite quantum systems in arbitrary
dimensions.

\noindent

\vskip0.2cm \textbf{Theorem 1:} \label{theorem1} \emph{For any density
matrix $\rho $, we can associate with it an EW defined as
\begin{equation}
W=Id-(\mathcal{R}^{-1}(U^{\ast }V^{T}))^{T},  \label{witness}
\end{equation}%
where $U,V$ are the unitary matrices that yield the singular value
decomposition (SVD) of $\mathcal{R}(\rho )$, i.e., $\mathcal{R}(\rho
)=U\Sigma V^{\dagger }.$}

\vskip0.2cm \noindent \textbf{Proof:} Using a result of matrix analysis (see
chapter $3$ of \cite{hornt}), we have%
\begin{align}
||\mathcal{R}(\rho )||& =\max \{|Tr(X^{\dagger }\mathcal{R}(\rho )Y)|:X\in
M_{m^{2},q},  \notag \\
& Y\in M_{n^{2},q},X^{\dagger }X=Id=Y^{\dagger }Y\},
\end{align}%
where $q=\min \{m^{2},n^{2}\}$. We thus find that the maximum value for $%
|TrX^{\dagger }\mathcal{R}(\rho )Y|$ occurs at $X=U$ and $Y=V$. This is
because
\begin{eqnarray*}
|Tr(U^{\dagger }\mathcal{R}(\rho )V)| &=&|Tr(U^{\dagger }U\Sigma V^{\dagger
}V)| \\
&=&|Tr\Sigma |=Tr\Sigma =||\mathcal{R}(\rho )||.
\end{eqnarray*}%
In the same way, for a separable state $\rho _{sep}$, $|Tr(X^{\dagger }%
\mathcal{R}(\rho _{sep})Y)|$ has its maximum value at $X=U^{\prime }$ and $%
Y=V^{\prime }$ where $\mathcal{R}(\rho _{sep})=U^{\prime }\Sigma ^{\prime }{%
V^{\prime }}^{\dagger }$ is the SVD of $R(\rho _{sep})$. From the
realignment criterion for separability we have $||\mathcal{R}(\rho
_{sep})||\leq 1$, thus
\begin{eqnarray*}
&&|Tr(U^{\dagger }\mathcal{R}(\rho _{sep})V)| \\
&\leq &|Tr({U^{\prime }}^{\dagger }\mathcal{R}(\rho _{sep})V^{\prime })|=||%
\mathcal{R}(\rho _{sep})||\leq 1.
\end{eqnarray*}%
Since $\mathcal{R}(\rho ^{\prime })$ is just a rearrangement of entries in $%
\rho ^{\prime }$, we find by direct observation that $W_{2}=(\mathcal{R}%
^{-1}(W_{1}^{T}))^{T}$ if we require $|Tr(W_{1}\mathcal{R}(\rho ^{\prime
}))|=|Tr(W_{2}\rho ^{^{\prime }})|$ to hold for all $\rho ^{\prime }$. Here $%
\mathcal{R}^{-1}(W_{1}^{T})$ means the inverse of $\mathcal{R}$, realigning
the entries of $W_{1}^{T}$ according to Eq. (\ref{realign}). Letting $%
W_{1}=VU^{\dagger }$, since $Tr(W\rho _{sep})=1-Tr(W_{2}\rho _{sep})\geq
1-|Tr(W_{2}\rho _{sep})|=1-|Tr(W_{1}\mathcal{R}(\rho _{sep}))|\geq 0$, we
have the EW
\begin{eqnarray}
W=Id-W_{2} &=&Id-(\mathcal{R}^{-1}(W_{1}^{T}))^{T}  \notag \\
&=&Id-(\mathcal{R}^{-1}(U^{\ast }V^{T}))^{T}.
\end{eqnarray}%
\hfill \rule{1ex}{1ex}

\vskip0.2cm Whenever we have an entangled state $\rho$ which can be detected
by the realignment criterion, i.e., $||\mathcal{R}(\rho)||>1$, it can also
be detected by the EW in Theorem 1, since $Tr(W\rho)=1-Tr(VU^{\dagger}%
\mathcal{R}(\rho))=1-||\mathcal{R}(\rho)||<0$. It should be remarked that
for an $m\times m$ system, we have $(\mathcal{R}^{-1}(U^{\ast}V^{T}))^{T}%
\equiv\mathcal{R}(VU^{\dagger})$ which is a simpler expression for practical
operation by direct observation.

As for the $PPT$ criterion, we can also have a universal construction for
EWs as follows:

\vskip0.2cm \noindent \textbf{Theorem 2:} \emph{For any density matrix $\rho$%
, we can associate with it another EW defined as
\begin{equation}
W=Id-(VU^{\dagger})^{T_{A}},  \label{witness2}
\end{equation}
where $U,V$ are unitary matrices that yield the SVD of $\rho^{T_{A}}$, i.e.,
$\rho^{T_{A}}=U\Sigma V^{\dagger}.$}

\vskip0.2cm \noindent\textbf{Proof:} For any separable density matrix $%
\rho_{sep}$, we have $||\rho_{sep}^{T_{A}}||=1$ due to positivity of $%
\rho_{sep}^{T_{A}}$. Similar to the procedure in the proof of Theorem 1, we
have
\begin{eqnarray}
Tr(W\rho _{sep})&=&1-Tr((VU^{\dagger})^{T_{A}}\rho_{sep})  \notag \\
&\geq& 1-|Tr((VU^{\dagger})^{T_{A} }\rho_{sep})|  \notag \\
&=&1-|Tr(VU^{\dagger}\rho_{sep}^{T_{A}})|  \notag \\
&\geq& 1-||\rho_{sep}^{T_{A}}||=0.
\end{eqnarray}
Thus $W$ is an EW. \hfill\rule{1ex}{1ex}

\vskip0.2cm Whenever we have an entangled state $\rho$ which can be detected
by the $PPT$ criterion, i.e., $||\rho^{T_{A}}||>1$, it can also be detected
by the EW in Theorem 2, since $Tr(W\rho)=1-Tr((VU^{\dagger})^{T_{A}}%
\rho)=1-||\rho^{T_{A}}||<0$.

\subsection{Optimization of EWs and universal construction of PMs}

\label{sec2.5}

The universal EWs that we have constructed from Theorem 1 and Theorem 2 are
no weaker than the $PPT$ criterion and the realignment criterion. We
anticipate that better tests should exist. Motivated by the idea developed
in Ref. \cite{lkch00}, we now derive a better witness $W^{\prime }$ from
Theorems 1 and 2:
\begin{equation}
W^{\prime }=W-\epsilon Id,  \label{optimize}
\end{equation}%
where $\epsilon =\min Tr(W\rho _{A}\otimes \rho _{B})$ for all possible pure
states $\rho _{A}$ and $\rho _{B}.$ We observe that $Tr(W^{\prime }\rho
_{sep})=Tr(W\rho _{sep})-\epsilon \geq 0$ since $Tr(W\sum_{i}p_{i}\rho
_{i}^{A}\otimes \rho _{i}^{B})=\sum_{i}p_{i}Tr(W\rho _{i}^{A}\otimes \rho
_{i}^{B})\geq \sum_{i}p_{i}\epsilon =\epsilon $. Thus $W^{\prime }$ is a
reasonable EW.

Let us see how the minimum of $Tr(W\rho _{A}\otimes \rho _{B})$ for a given $%
W$ is obtained. Partitioning $W$ to be an $m\times m$ block matrix $W_{i,j}$
$(i,j=1,\ldots m)$ with block size $n\times n$, we have%
\begin{eqnarray}
\epsilon  &=&\min Tr(W\rho _{A}\otimes \rho _{B})  \notag \\
&=&\min Tr\left\{ \left[ \sum\nolimits_{i,j}W_{i,j}(\rho _{A})_{ji}\right]
\rho _{B}\right\} .
\end{eqnarray}%
Here $W_{i,j}$ is an $n\times n$ matrix while $(\rho _{A})_{ji}$ is a single
entry of $\rho _{A}$. Using a known result of matrix analysis:
\begin{equation*}
\lambda _{\min }(G)=\min Tr(U^{\dagger }GU)=\min Tr(GUU^{\dagger }),
\end{equation*}%
where $U^{\dagger }U=1$ and $\lambda _{\min }(G)$ is the minimum eigenvalue
of $G$, we deduce that $\epsilon $ is in fact the minimum eigenvalue of $%
G=\sum\nolimits_{i,j}W_{i,j}(\rho _{A})_{ji})$. Thus the problem changes to
that of finding $\lambda _{\min }(G)$ for all possible $\rho _{A}$, which
can be done with common numerical optimization software. This is much
simpler for the low dimensional cases of the first subsystem such as $m=2,3$.

However, the ability of the optimized $W^{\prime }$ to detect entanglement
is still not ideal, and some stronger entanglement tests than the EWs should
be found. The Jamio\l kowski isomorphism Eq. (\ref{isomor}) between a PM and
an EW serves us a good candidate. A positive map can in fact detect more
entanglement than its corresponding witness operator \cite{3hreview}.
According to Eq. (\ref{isomor}) the positive map $\Lambda $ corresponding to
a given $W$ is of the form
\begin{equation}
\Lambda (|i\rangle \langle j|)=\langle i|W|j\rangle ,  \label{pomap}
\end{equation}%
where $\{|i\rangle \}_{i=1}^{m}$ is an orthogonal basis for $\mathcal{H}^{A}$%
.

\vskip0.2cm Using Theorems 1 and 2 we now have a universal EW construction
to detect entanglement for a given quantum state, which is not weaker than
the two criteria for separability. We can also detect all other possible
entangled states (in particular the same class of states) using the
constructed EW. Moreover, we can use the optimizing procedure of Eq. (\ref%
{optimize}) to get a better witness, while the best detection can be
obtained with the positive map $\Lambda $ given by Eq. (\ref{pomap}). We say
that our constructions are universal in the sense that we can, in principle,
find an EW and a PM to detect entanglement in most of the entangled states
(especially the bound entangled states). If we have some prior knowledge of
a quantum state, we can even detect its entanglement with just a single EW
and a PM from the constructions of Theorem 1 and 2, without having to apply
full quantum state tomography \cite{tomog1}. The $2\times 2$ Werner state
\cite{VoPRA01}) is such an example as shown in the following section. To our
surprise, we find that even for some separable states we can still obtain
good EWs to detect many entangled states. Additional examples are given
below for bound entangled states to depict this character. The constructed
EWs, their optimized versions and corresponding PM are always more powerful
than the $PPT$ criterion and the realignment criterion.

\section{Application of EWs and PMs to entangled states}

\label{sec3} In this section we give several typical examples to display the
power of our constructions to distinguish entangled states, in particular,
the bound entangled states from separable states. Example 1 is a simple $%
2\times 2$ Werner state, which is to show that we can still obtain good EWs
and PMs even with a separable states. Since the $PPT$ criterion is strong
enough to detect all non-$PPT$ entangled states we shall just examine PPTES
in the following two examples. We have tested most of the bipartite BESs in
the literature \cite{UPB,hPLA97,bruss} and found that the optimized EW $%
W^{\prime }$ and corresponding PM are always much more powerful than the
realignment criterion.

\vskip0.2cm \noindent \emph{Example 1:} $2\times 2$ Werner state.

Consider the family of two-dimensional Werner states \cite{werner89},
\begin{equation}
\rho =\frac{1}{6}\big(\left( 2-f\right) \mathbb{I}_{4}+(2f-1)V\big),
\label{werner}
\end{equation}%
where $-1\leq f\leq 1$, $V(\alpha \otimes \beta )=\beta \otimes \alpha $,
the operator $V$ has the representation $V=\sum_{i,j=0}^{1}|ij\rangle
\langle ji|$, and $\rho $ is inseparable if and only if $-1\leq f<0$. As is
well known, the entanglement in a $2\times 2$ Werner state can be detected
completely using the \emph{PPT} criterion and the realignment criteria. It
is straightforward to verify that we can obtain a single EW
\begin{equation}
W=%
\begin{pmatrix}
1 & 0 & 0 & 0 \\
0 & 0 & 1 & 0 \\
0 & 1 & 0 & 0 \\
0 & 0 & 0 & 1%
\end{pmatrix}%
,
\end{equation}%
from Theorem 1 whenever $-1\leq f<1/2$. From $Tr(W\rho )=f<0$, we see that
this EW and corresponding PM can detect all of the entangled $2\times 2$
Werner states. This is a surprising result since we still obtain a good
witness operator associated with a separable state ($0\leq f<1/2$) to detect
the whole family of Werner states.

\vskip0.2cm \noindent \emph{Example 2}: $3\times3$ BES constructed from
unextendible product bases

\noindent In Ref. \cite{UPB}, Bennett et al introduced a $3\times 3$
inseparable BES from the following bases:
\begin{align}
{|\psi _{0}\rangle }& ={\frac{1}{\sqrt{2}}}{|0\rangle }({|0\rangle }-{\
|1\rangle }),\ \ \ \ {|\psi _{1}\rangle }={\frac{1}{\sqrt{2}}}({|0\rangle }-{%
\ |1\rangle }){|2\rangle },  \notag \\
{|\psi _{2}\rangle }& ={\frac{1}{\sqrt{2}}|2\rangle }({|1\rangle }-{\
|2\rangle }),\text{ \ \ \ }{|\psi _{3}\rangle }={\frac{1}{\sqrt{2}}}({%
|1\rangle }-{|2\rangle }){|0\rangle },  \notag \\
{|\psi _{4}\rangle }& ={\frac{1}{3}}({|0\rangle }+{|1\rangle }+{|2\rangle )}(%
{|0\rangle }+{|1\rangle }+{|2\rangle )},
\end{align}%
from which a bound entangled density matrix can be constructed as
\begin{equation}
\rho =\frac{1}{4}(Id-\sum_{i=0}^{4}{|\psi _{i}\rangle \langle \psi _{i}|}).
\end{equation}%
Consider a mixture of this state and the maximally mixed state $\rho
_{p}=p\rho +(1-p)Id/9$. Direct calculation using the realignment criterion
gives $p>88.97$ $\%$ where $\rho _{p}$ still has entanglement. This is much
stronger than the optimal witness given in Ref. \cite{terhal00,guhne} where
they give $p>94.88$ $\%$ for the existence of entanglement in $\rho _{p}$.
Using Theorem 1 for $\rho $, we obtain an EW (not optimized) and further a
positive map to detect entanglement for $\rho _{p}$ when $p>88.41$ $\%$. A
surprising result is that we can still obtain a good EW to detect
entanglement in $\rho $ whenever we generate an EW from $\rho _{p}$ using
Theorem 1 for all values of $0<p\leq 1$. Numerical calculation shows that
the best PM is the one corresponding to the EW generated from $\rho _{p}$
when $p\doteq 0.3$, which is, to a large degree, a separable state (we have
no operational separability criterion to guarantee this but we can estimate
that it is so). The best PM detects entanglement for $\rho _{p}$ when $%
p>87.44$ $\%$, which is, to our knowledge, the strongest test up to now.

\vskip0.2cm \noindent \emph{Example 3}: $3\times3$ chessboard BES

\noindent Bru{\ss } and Peres constructed a seven parameter family of PPTES
in~Ref. \cite{bruss}. Using the above mentioned constructions we perform a
systematic test for these states. Choosing a state with a relatively large $%
||\mathcal{R}(\rho )||=1.164$ in the family and constructing a PM from the
EW corresponding to $\rho $ we can detect about $9.48$ $\%$ of $10^{4}$
randomly chosen density matrix $\sigma $ satisfying $\sigma =\sigma ^{T_{A}}$%
. Half of these detected BESs cannot be detected by the realignment
criterion, and it should also be noted that we only used one PM here. For
every state in this family (including those that cannot be detected by the
realignment criterion) we can almost always obtain an EW and a PM which can
detect some BESs in this family, many of which cannot be recognized by the
realignment criterion by a direct numerical calculation.

\vskip0.2cm Actually, we have a third choice in constructing an EW with
\begin{equation*}
W=\epsilon Id-\rho ,
\end{equation*}%
where $\epsilon =\max Tr(\rho (\rho _{A}\otimes \rho _{B}))$ for a given
density matrix $\rho $, as first proposed in Ref. \cite{lbck00}. We can
calculate $\epsilon $ following the same procedure as for optimizing the EW,
and find that $\epsilon $ is actually the maximum eigenvalue of $%
G=\sum\nolimits_{i,j}W_{i,j}(\rho _{A})_{ji})$ for all possible $\rho _{A}$.
The optimal witness given in Ref. \cite{guhne} for states constructed from
unextendible product bases is in fact equivalent to this method.

Since some PMs can detect BES they cannot be decomposed to the form $\Lambda
_{1}+T\circ \Lambda _{1}$ where $\Lambda _{1}$ and $\Lambda _{2}$ are
completely positive maps, and $T$ is the standard transposition \cite%
{3hPLA223}. Thus our construction gives a universal method to find the
indecomposable positive linear map in any dimension.

\section{two indecomposable positive maps}

\label{sec4}

\begin{figure}[tbh]
\begin{center}
\resizebox{8cm}{!}{\includegraphics{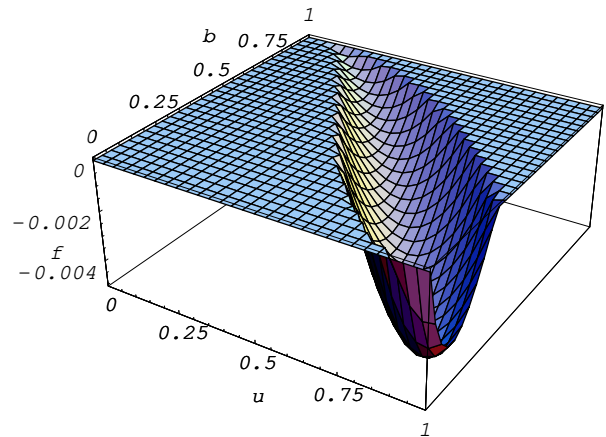}}
\end{center}
\caption{Detection of a Horodecki $2\times 4$ bound entangled state with the
indecomposable positive map $\Lambda $ of Eq.(\protect\ref{4x2pm}).}
\label{fig1}
\end{figure}

Here we also show that two indecomposable positive maps, which were first
given in Ref. \cite{tang86}, can systematically detect the $2\times 4$ BES
given by Horodecki \cite{hPLA97}. One map $\Lambda :M_{4}\longrightarrow
M_{2}$ is defined as $\Lambda :[a_{ij}]_{i,j=1}^{4}\longrightarrow $%
\begin{equation}
\left(
\begin{array}{c|c}
\begin{array}{r}
(1-\varepsilon )a_{11}+a_{22} \\
+2a_{33}+a_{44}%
\end{array}
&
\begin{array}{r}
-2a_{23}-2a_{34} \\
+ua_{31}-a_{12}%
\end{array}
\\ \hline
\begin{array}{r}
-2a_{32}-2a_{43} \\
+ua_{13}-a_{21}%
\end{array}
&
\begin{array}{r}
u^{2}a_{11}-ua_{14}+2a_{22} \\
-ua_{41}+a_{44}%
\end{array}%
\end{array}%
\right) ,  \label{4x2pm}
\end{equation}%
where $0<u<1$ and $0<\varepsilon \leq u^{2}/6$. The density matrix $\rho $
of the $2\times 4$ BES in Ref. \cite{hPLA97} is real and symmetric, and has
the form:
\begin{equation}
\rho ={\frac{1}{7b+1}}\left[
\begin{array}{ccccccccc}
b & 0 & 0 & 0 & 0 & b & 0 & 0 &  \\
0 & b & 0 & 0 & 0 & 0 & b & 0 &  \\
0 & 0 & b & 0 & 0 & 0 & 0 & b &  \\
0 & 0 & 0 & b & 0 & 0 & 0 & 0 &  \\
0 & 0 & 0 & 0 & {\frac{1+b}{2}} & 0 & 0 & {\frac{\sqrt{1-b^{2}}}{2}} &  \\
b & 0 & 0 & 0 & 0 & b & 0 & 0 &  \\
0 & b & 0 & 0 & 0 & 0 & b & 0 &  \\
0 & 0 & b & 0 & {\frac{\sqrt{1-b^{2}}}{2}} & 0 & 0 & {\frac{1+b}{2}} &
\end{array}%
\right] ,
\end{equation}%
where $0<b<1$. Assuming $\varepsilon =u^{2}/6$, we see that $(Id_{A}\otimes
\Lambda )\rho $ can detect all the entanglement in $\rho $ for $0<b<1$, as
shown in Fig. \ref{fig1}, where we have plotted $f=\min \{0,\lambda _{\min }%
\left[ (Id_{A}\otimes \Lambda )\rho \right] \}$, and $\lambda _{\min }$
means the minimum eigenvalue. It is straightforward to verify that the dual
map $\Lambda ^{\prime }:M_{2}\longrightarrow M_{4}$ to $\Lambda $ in Ref.
\cite{tang86} can also detect $\rho $ by the action of $(\Lambda ^{\prime
}\otimes Id_{B})\rho $. If we assume $\rho _{p}=p\rho +(1-p)Id/8$, the PM $%
\Lambda $ gives $p>99.26$ $\%$ for existence of entanglement in $\rho _{p}$
with $u=0.849$ and $b=0.218$, which is a stronger test than the PM
constructed from an optimal EW shown in Ref. \cite{lkch00} to give $p>99.65$
$\%$. For any $2\times N$ or $4\times N$ system, the maps $\Lambda $ and $%
\Lambda ^{\prime }$ are expected to give a very strong test for recognizing
BES.

\section{Conclusion}

\label{sec5}

Summarizing, we have presented a universal construction for entanglement
witnesses and positive maps that can detect entanglement systematically and
operationally. They are stronger than both the $PPT$ and the realignment
criteria, and provide a powerful method to detect entanglement since
entanglement witnesses are physical observables and may be measured locally
\cite{guhne}. The construction gives us a new subtle way to enhance
significantly our ability to detect entanglement beyond previously known
methods. Even when associated with some separable states, the construction
gives EWs and PMs to identify entanglement. If we have some prior knowledge
of a quantum state, we can even detect its entanglement with just a single
EW and a PM. In addition, our construction gives a systematic way to obtain
positive but non-CP maps, which may also be of interest to the mathematics
community. Moreover, we find that two types of positive maps can detect
completely a $2\times 4$ bound entangled state and promise to give a very
strong test for any $2\times N$ and $4\times N$ systems. We hope that our
method can shed some light on the final solution of the separability
problem, as well as motivate new interdisciplinary studies connected with
mathematics.

\section*{ACKNOWLEDGEMENTS}
The authors would like to thank Shao-Ming Fei and
Heng Fan for stimulating discussions. K.C. is grateful to Guozhen Yang for
his continuous encouragement. This work was supported by the Chinese Academy
of Sciences, the National Program for Fundamental Research, the National
Natural Science Foundation of China, and the China Postdoctoral Science
Foundation.

\end{document}